\documentclass[preprints,article,accept,moreauthors,pdftex]{Definitions/mdpi}

\usepackage{isotope}

\usepackage{soul}
\usepackage{xcolor}

\sethlcolor{white} 

\firstpage{1}
\pubvolume{1}
\issuenum{1}
\articlenumber{0}
\pubyear{2025}
\copyrightyear{2025}
\externaleditor{Ann Morrison and Vincent Breton}
\datereceived{1 November 2024}
\daterevised{6 February 2025} 
\dateaccepted{12 February 2025}
\datepublished{}
\hreflink{https://doi.org/} 
\Title{Optimizing 
Sensor Data Interpretation via Hybrid Parametric~Bootstrapping}

\TitleCitation{Optimizing Sensor Data Interpretation via Hybrid Parametric Bootstrapping}

\Author{\hl{Victor V. Golovko \orcidA{}} 
}

\AuthorNames{Victor V. Golovko}

\AuthorCitation{Golovko, V.V.}

\address[1]{Canadian Nuclear \hl{Laboratories,} 
 286 Plant Road, Chalk River, ON K0J~1J0, Canada; \hl{victor.golovko@cnl.ca} 
}

\abstract{
The Chalk River Laboratories (CRL) site in Ontario, Canada, has long been a hub for nuclear research, which has resulted in the accumulation of legacy nuclear waste, including radioactive materials such as uranium, plutonium, and other radionuclides. Effective management of this legacy requires precise contamination and risk assessments, with a particular focus on the concentration levels of fissile materials such as $^{235}$U. These assessments are essential for maintaining nuclear criticality safety. This study estimates the upper bounds of $^{235}$U concentrations. We investigated the use of a hybrid parametric bootstrapping method and robust statistical techniques to analyze datasets with outliers, then compared these outcomes with those derived from nonparametric bootstrapping. This study underscores the significance of measuring $^{235}$U for ensuring safety, conducting environmental monitoring, and adhering to regulatory compliance requirements at nuclear legacy sites. We used publicly accessible $^{235}$U data from the Eastern Desert of Egypt to demonstrate the application of these statistical methods to small datasets, providing reliable upper limit estimates that are vital for remediation and decommissioning efforts. This method seeks to enhance the interpretation of sensor data, ultimately supporting safer nuclear waste management practices at legacy sites such as CRL.
}


\begin{document}

\section{
Introduction
}

\begin{figure}[t]
	\centering
	\includegraphics[width=\textwidth, trim=0 0 0 10pt, clip]{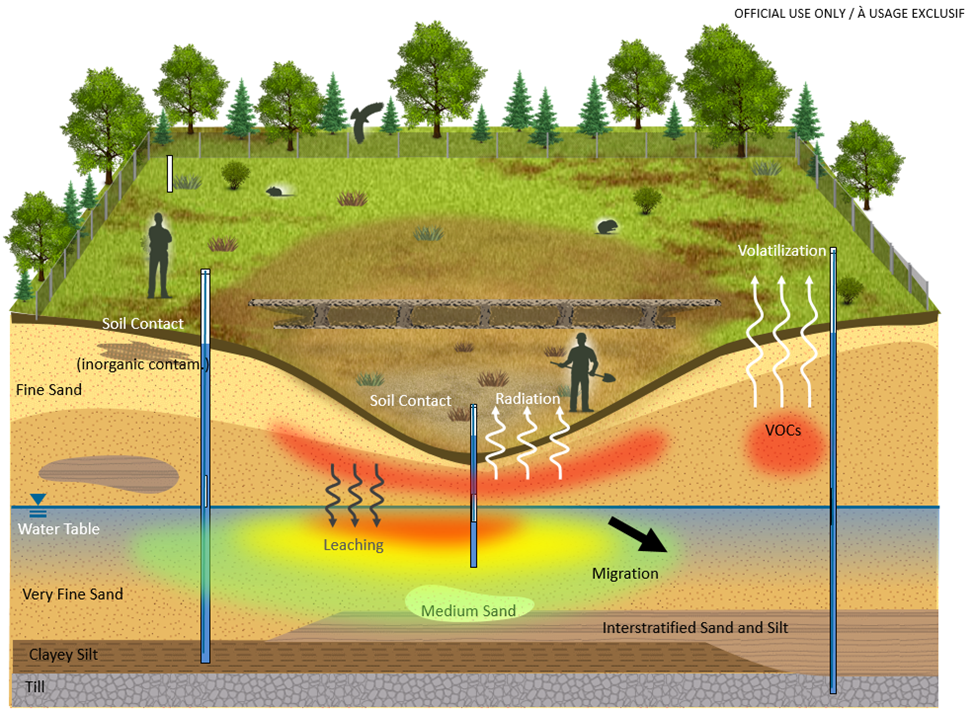} \\
	\caption{\hl{A conceptual} 
 site model for a nuclear waste area at Chalk River~Laboratories. }
	\label{fig:CSM_legasy}
\end{figure}

The Chalk River Laboratories (CRL) site in Ontario, Canada is notable for its extensive history of nuclear research and development, including the operation of research reactors and related facilities. Several legacy waste sites at the CRL have been identified. These sites store, treat, or~dispose of radioactive and hazardous materials. The~conceptual site model (CSM) for these waste sites, shown in Figure~\ref{fig:CSM_legasy}, combines historical data, the~specific environmental conditions of the site, and~the expected behavior of the contaminants. This model is crucial for directing remediation efforts and conducting effective risk~assessments.

The legacy waste sites at CRL are mainly connected to nuclear research activities, which encompass reactor operations, radiochemical laboratories, and~waste handling methods. These sites contain waste materials such as fission products, actinides (including uranium, plutonium, and~americium), and~other radionuclides and hazardous chemicals. In~the past, disposal methods included onsite storing of waste in specially designed pits, trenches, and~infiltration areas, some of which led to underground~contamination.

Before engaging in any tasks involving the handling, storage, or~remediation of facilities that house fissionable materials, it is essential to develop a robust criticality safety strategy. This study offers a concise outline of the fundamental components of this strategy, derived from an analysis of the specific nuclear materials present at a typical nuclear legacy waste site at the~CRL.

A typical nuclear legacy waste site contains radioactive substances such as uranium, plutonium, americium, cesium, strontium, and~technetium, which are often found in soil, water, and~sediments. There are also nonradioactive materials present, which include volatile organic compounds (VOCs), heavy metals, and~chemical solvents. These hazardous materials can disperse in several ways: they may wash into nearby water sources such as the Ottawa River due to surface runoff, seep into groundwater, migrate through soil, or~become airborne as dust. In~addition, VOCs pose a special risk because they can evaporate, turn into gases, and~pollute the~air.

Humans, including workers and the general public, as~well as the environment, including wildlife and aquatic life, are at risk of harmful exposure. Contaminants or VOCs present in the air can affect individuals through inhalation, ingestion, or~direct contact. Similarly, ecosystems can suffer from pollutants present in contaminated soil, water, and~air.

The CRL actively monitors and manages different environmental factors, notably radioactivity; in certain areas, specific actions are prioritized in order to meet safety and environmental guidelines. Remediation techniques include methods such as capping, groundwater treatment, continuous monitoring over extended periods, and~various other~strategies.

During the remediation and decommissioning processes, it is crucial to maintain nuclear criticality safety in order to prevent the dangerous buildup of fissile materials. This task involves evaluating the risks associated with criticality and establishing safety measures to prevent the formation of a critical mass during~operations.

Disclosing specific activity values for fissile materials in nuclear legacy waste is not recommended because they can be misused. To~responsibly manage such waste, access to sensitive data must be strictly controlled. Instead, we used publicly available data to perform advanced statistical analyses in order to develop safer remediation methods. These methods are adaptable to various~datasets.

This study examined the concentrations of $^{235}$U found in granite samples from the Eastern Desert of Egypt, as~reported in publicly accessible data~\cite{harb2008concentration}; the selected dataset was used because of its extensive element inclusion, which makes it ideal for employing a nonparametric bootstrapping technique. Measurements of $^{235}$U activity concentrations were measured using a high-purity Germanium (HPGe) spectrometer. This study also considers the variable presence of naturally occurring radioactive materials (NORM) in sedimentary rocks, which can indicate the sediment's origin and depositional context along with the geological development of the surrounding~basin.

This project focuses on determining the maximum concentration levels of $^{235}$U in contaminated soils from nuclear legacy sites. Given the security constraints, granite was used to simulate these concentrations. A~robust statistical approach was applied to effectively manage datasets with potential outliers. Furthermore, this research demonstrates the hybrid parametric bootstrapping technique for estimating the upper concentration limits from small datasets with fewer than ten elements and contrasts these findings with results acquired through nonparametric bootstrapping, ensuring the accuracy and reliability of the~results.

In nuclear remediation, the~term ``upper limit of $^{235}$U concentrations'' generally pertains to the upper confidence level (UCL). The~UCL is a statistical method used to verify that radioactive contamination levels comply with regulatory standards, especially during radiation surveys. The~proposed method sets a statistical upper boundary for the contamination levels, which helps to ensure that the remediation efforts are aligned with safety~requirements.

This method addresses uncertainties in the measurement data in order to ensure that the reported contamination levels remain below the regulatory limits within a given confidence level. The~UCL technique aids in making more precise decisions about site release criteria and the safe handling or disposal of radioactive~materials.

Measuring the concentration of $^{235}$U at sites with a nuclear history is crucial for a number of reasons. First, for~safety reasons, precise measurements of $^{235}$U are vital to prevent unintended criticality incidents and ensure nuclear safety. Second, from~an environmental perspective, observing $^{235}$U helps in evaluating contamination levels and distinguishing between natural uranium and uranium resulting from human activities. Third, during~decommissioning and site cleanup, $^{235}$U measurements help to characterize nuclear waste, assisting in the correct classification of waste for disposal. Lastly, ensuring regulatory compliance requires verifying that $^{235}$U levels adhere to the safety standards necessary for license termination and environmental impact evaluations. These measurements empower site managers to make well informed decisions regarding the safety, remediation, and~sustainable management of the~site.

\section{{Materials and Methods} 
 }

When analyzing the data, it is crucial to confirm that the sample size adequately represents the location or exposure area, particularly if the concentrations vary greatly. The~sampling methods play a key role; at many hazardous waste sites, they typically focus on zones suspected of being contaminated. After~validating the data, an~exploratory analysis is required to identify any outliers or significant `nondetect' values, both of which can affect the statistical outcomes. A~`nondetect' in data samples means that the substance being tested for is present at a level too low for the equipment or method in question to reliably measure or~detect.

Outliers are data points that deviate significantly from the majority, often because they have much higher values. In~contaminated site data, the~concentration measurements are typically skewed, highlighting a few high values that indicate possible local hot spots. Although~these large values are outliers, they should not be dismissed without careful analysis, as~they may represent genuine variations in the contamination~levels.

Robust statistical methods are used to reduce the impact that outliers have on estimates such as the mean or standard deviation. These methods highlight the importance of separately investigating outliers. Steiner's most frequent value (MFV) statistic is effective for identifying the central tendency of data while minimizing the influence of outliers. When paired with a nonparametric bootstrapping method~\cite{efron1994introduction} which does not rely on any specific data distribution, the~MFV helps to estimate both the variability and central tendency even in the presence of outliers. This combination also enables calculation of the upper confidence limit in the assessment of nuclear legacy contaminated sites, providing a more reliable~analysis.

\subsection{The Most Frequent Value Method}
\label{subs:MFV}

The MFV method is used in geophysics to make data estimation more efficient, enabling geophysicists to gather precise geological information while reducing expenses in line with economic efficiency principles in the field~\cite{CsernyakSteiner1983}. Furthermore, the~MFV approach effectively addresses missing data by using the most frequent values to determine the cluster centers, thereby lessening the impact of incomplete data in geophysical datasets~\cite{MOHAMMED2024102594}. \hl{Appendix B} 
 of Golovko (2024)~\cite{GOLOVKO2024143910} provides a short summary of the MFV method for the dataset analysis. Additionally, the~codes for calculating the MFV and its variance for symmetrical data are available in the referenced repository~\cite{GolovkoDataSimplifiedEff2024}.

The MFV method, first introduced by Steiner~\cite{steiner1973most, steinerMostFrequentValue1991, steinerOptimumMethodsStatistics1997}, offers a reliable technique for identifying the most frequent value in a dataset ($x_1, \dots, x_i, \dots, x_N$). This approach involves repeatedly computing the MFV ($M$) and the scale parameter $\varepsilon$, known as dihesion. The~iterative calculation for determining $M_{j+1}$ is specified as \hl{follows:} 
\begin{equation} M_{j+1} = \frac{ \sum_{i=1}^{N} x_i \cdot \frac{1}{\varepsilon^2_j + \left( x_i - M_j \right)^2 } }{ \sum_{i=1}^{N} \frac{1}{\varepsilon^2_j + \left( x_i - M_j \right)^2} }. \label{Eq:M}
\end{equation}

The scale parameter $\varepsilon_j$, which is also referred to as dihesion, is updated in the following~steps:
\begin{equation} \varepsilon^2_{j+1} = 3 \cdot \frac{ \sum_{i=1}^{N} \frac{( x_i - M_j)^2 }{ \left( \varepsilon^2_j + \left( x_i - M_j \right)^2 \right)^2 } }{ \sum_{i=1}^{N} \frac{ 1 }{ \left( \varepsilon^2_j + \left( x_i - M_j \right)^2 \right)^2 } }. \label{Eq:eps}
\end{equation}

The iterative process begins by establishing the initial values for $M_{(0)}$ and $\varepsilon_{(0)}$. For~$M_{(0)}$, the~starting value is the median, as~described by Hajagos and Steiner (1992)~\cite{HajagosSteiner1992}; regarding $\varepsilon_{(0)}$, Hajagos (1980)~\cite{Hajagos1980} proposed an efficient method to calculate it:
\begin{equation} \varepsilon_{(0)} = \frac{\sqrt{3}}{2} \cdot (x_{\text{max}} - x_{\text{min}})
\end{equation}
where $x_{\text{max}}$ and $x_{\text{min}}$ stand for the maximum and minimum values in the dataset, respectively. This initial guess facilitates rapid convergence in the iterative calculation of the~MFV.

To accurately compute the most frequent value (MFV, denoted as $M$) and the scale parameter ($\varepsilon$), it is {necessary} to solve both Equations~(\ref{Eq:M}) and (\ref{Eq:eps}) iteratively until they converge. This process can require substantial computational effort, particularly when working with large datasets; however, it provides a reliable estimate of the most frequent value. For~practical applications of this method, refer to the works of Zhang~et~al. (2022, 2024)~\cite{zhang2022mfv,zhang2024most} and Golovko (2023, 2024)~\cite{golovkoApplicationMostFrequent2023,golovko2023unveiling}. {The MFV method has been explored in various practical contexts}~\cite{szaboMostFrequentValuebased2018,NaumahDobroka2019,dobrokaMFVbasedImageProcessing2021a,kilikHistogrambasedWeightedMedian2021b,szabo2023robust,Tolneretal2023,SZUCS2024130693,Tolner2024,MarashlyDobroka2024}.

For symmetrical distributions, Csernyak and Steiner~\cite{CsernyakSteiner1983} offer a simplified formula to compute the variance $\sigma_{M_j}$ of the MFV at each iteration. The~formula is provided by
\begin{equation}
	\sigma_{M_j} = \frac{1}{\sqrt{\sum_{i=1}^{N} \frac{1}{\varepsilon_j^2 + \left( x_i - M_j \right)^2}}}. \label{Eq:sMFV}
\end{equation}
\hl{In the above} 
 equation, $\varepsilon_j$ represents the dihesion factor, while each term in the summation represents the contribution of an individual data point in the dataset. This expression helps to calculate the variance, which is a critical measure for assessing the precision of $M$ in symmetric~distributions.

A recent study aimed at improving sustainable nuclear remediation practices introduced and verified new techniques for measuring the photopeak efficiency of thallium-activated sodium iodide scintillation detectors. Additionally, the~MFV method has been effectively applied to assess the stability of the active sensors, greatly improving the ability to screen radioactive isotopes at polluted sites~\cite{GOLOVKO2024143910, GolovkoDataSimplifiedEff2024}. In~another study that employed advanced statistical techniques, the~MFV method together with hybrid parametric bootstrap (HPB) analysis yielded more precise estimations of the $^{97}$Ru half-life and the specific activity of $^{39}$Ar measured in underground experiments, with~significantly tightened confidence intervals (CI) and lowered uncertainties~\cite{golovkoEstimation97RuHalfLife2024, golovkoDataEstimation97Ru2024}.

\subsection{Bootstrapping Method}\label{subs:boot}

To understand the uncertainty or variability of a statistical model or specific measure such as the MFV, researchers use a method called traditional bootstrapping, as~highlighted in Davison's work~\cite{davisonBootstrapMethodsTheir1997b}. This method involves repeatedly sampling from the existing dataset by selecting samples with replacement, termed bootstrap samples. By~analyzing these samples, researchers can create a distribution for the statistic of interest, such as the MFV, and~thereby for the confidence intervals, which helps to predict the range of possible values in order to make more informed decisions about the data or model in~question.

By combining the MFV method with bootstrapping, field data accuracy can be enhanced using existing historical data, thereby eliminating the need for additional experiments. In~practical applications, the~MFV method has been demonstrated to successfully manage outliers~\cite{golovkoApplicationMostFrequent2023,golovko2023unveiling}. To~provide an example of this, we include the uranium activity concentration dataset both with outliers and without.  As~the MFV approach focuses on minimizing information loss, integrating it with the bootstrapping techniques allows for more comprehensive and flexible averaging of the~results.

Bootstrapping offers advantages regardless of the presence of a defined probability model for the data~\cite{efron1994introduction,davisonBootstrapMethodsTheir1997b}. It enables the creation of a statistical distribution, which helps to make more precise inferences about the population or model under examination. The~traditional bootstrap method can be divided into two types, namely, nonparametric and parametric bootstrapping; nonparametric bootstrapping involves resampling directly from the original dataset, whereas parametric bootstrapping involves resampling from a model that fits the original data~\cite{puth2015variety}. This difference is crucial for understanding how the bootstrap method can be used in different~contexts.

The bootstrapping process is straightforward but requires significant computational power. Importantly, bootstrapping does not compensate for small sample sizes, and~does not create new data or fill in missing dataset values. Instead, it provides insights into how additional samples might behave if drawn from a population similar to the original sample~\cite{bruce2020practical}.

When dealing with a small dataset, the~confidence interval estimation can be improved by using a simple parametric bootstrap method. This method assumes that the data follow a specific distribution such as the normal distribution, and~does not account for individual uncertainties in the data measurements. It begins with a hypothesis about the data distribution, then uses estimated parameters such as the mean and variance from the original data to create new synthetic datasets. These synthetic datasets are then employed to recalculate the desired statistics, such as the mean or confidence interval. The~process is repeated multiple times to develop the distribution of the~estimator.

Examining the recalculated distribution of statistics helps us to understand the variability, potential skewness, and~reliability of the estimator's results, and~provides a basis for making statistical predictions based on the model's assumptions. This method is particularly useful when the data distribution is clearly defined and can be described using parameters, allowing for statistical tests and estimations that would be challenging if relying solely on direct sampling from the entire~population.

\section{
Description of the~Methods
}

This dataset details the NORM activity concentrations of $^{235}$U as documented by Harb (2008)~\cite{harb2008concentration}, consisting of 30 measurements reported in Bq per kilogram of granite from Egypt's Eastern Desert. These measurements, including their uncertainties indicated in parentheses, are as follows: 1.90 (0.18), 2.03 (0.19), 1.92 (0.18), 1.87 (0.17), 2.41 (0.22), \mbox{1.37 (0.13)}, 2.16 (0.20), 2.14 (0.20), 1.16 (0.11), 2.18 (0.20), 2.31 (0.21), 2.23 (0.21), 1.23 (0.11), 2.43 (0.22), 2.13 (0.20), 2.10 (0.19), 1.28 (0.12), 1.96 (0.18), 1.94 (0.18), 2.41 (0.22), 2.29 (0.21), 2.70 (0.25), 2.20 (0.20), 2.08 (0.19), 2.40 (0.22), 2.40 (0.22), 4.21 (0.39), 4.83 (0.44), 2.48 (0.23), and~2.27 (0.21). The~average (mean) of these concentrations is 2.23, with~a standard deviation of 0.73. The~dataset was assessed for any outliers using boxplot~analysis.

\begin{figure}[t]
	\includegraphics[width=\textwidth]{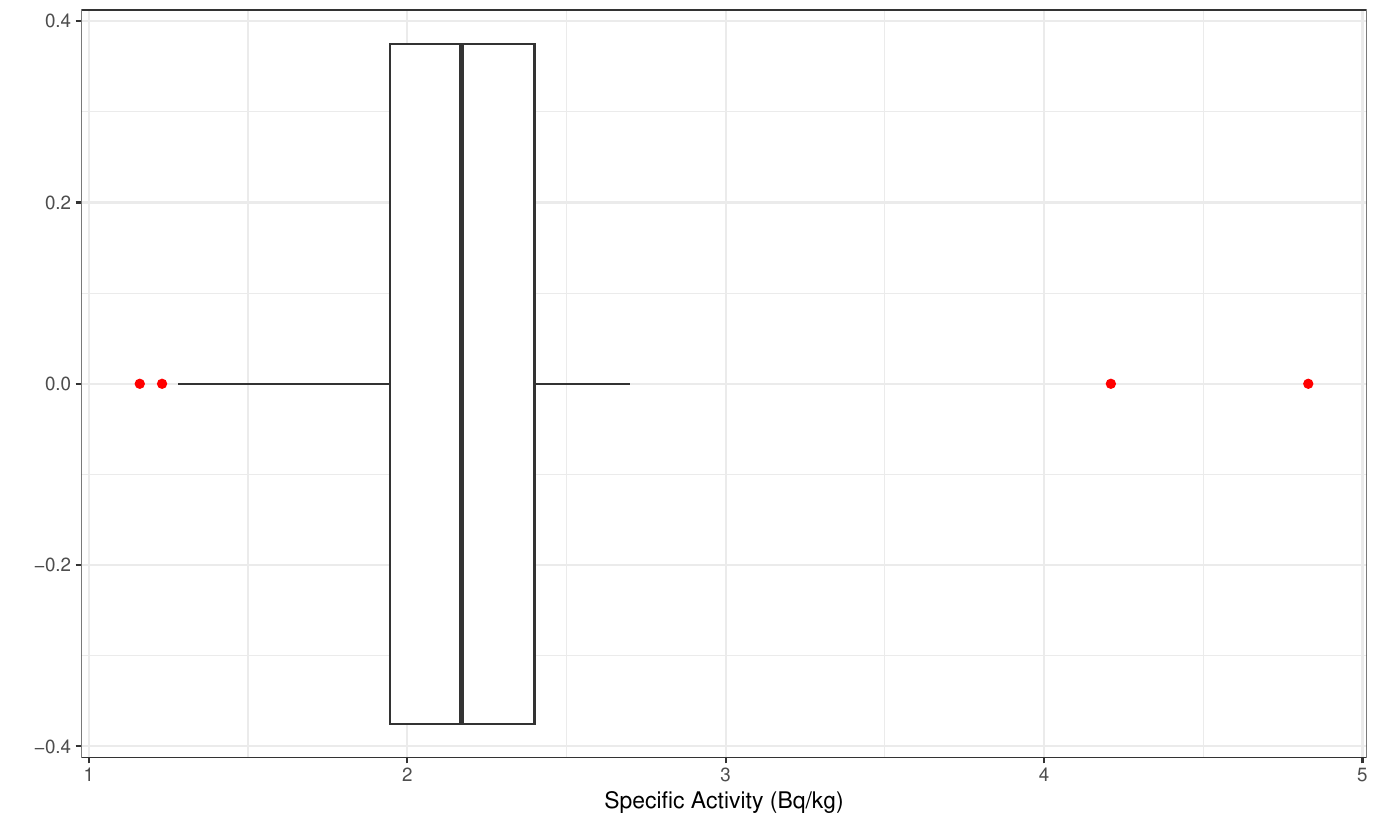}
	\caption{{\hl{Visualization} 
			of} \( ^{235}\text{U} \) {specific activity: median, IQR, and~outliers (red dots on the plot).} }
	\label{fig:U-235-boxplot}
\end{figure}

A boxplot (see {Figure~\ref{fig:U-235-boxplot})}
is a highly effective tool for spotting outliers in datasets. It provides a clear visual representation of the data's spread and variability, emphasizing any values that significantly deviate from the rest. The~central component of the plot (the box) highlights the interquartile range (IQR), which encompasses the central 50\% of the data points. Extending from the box are lines, called whiskers, which usually extend to 1.5 times the IQR above the third quartile and below the first quartile. Data points falling outside these whiskers are flagged as outliers, as~they deviate from the expected variability range. This technique's popularity originates from its nonparametric nature, meaning that it does not rely on a predefined data distribution; thus, it is adaptable across various datasets. Recognizing outliers is vital for statistical analyses, as~they can identify unusual patterns, potential data entry errors, or~important phenomena that warrant further exploration. In~this case, the~boxplot analysis identified the following outliers: 1.16 (0.11), 1.23 (0.11), 4.21 (0.39), and~4.83 (0.44).

\begin{figure}[t]
	
	\includegraphics[width=\textwidth]{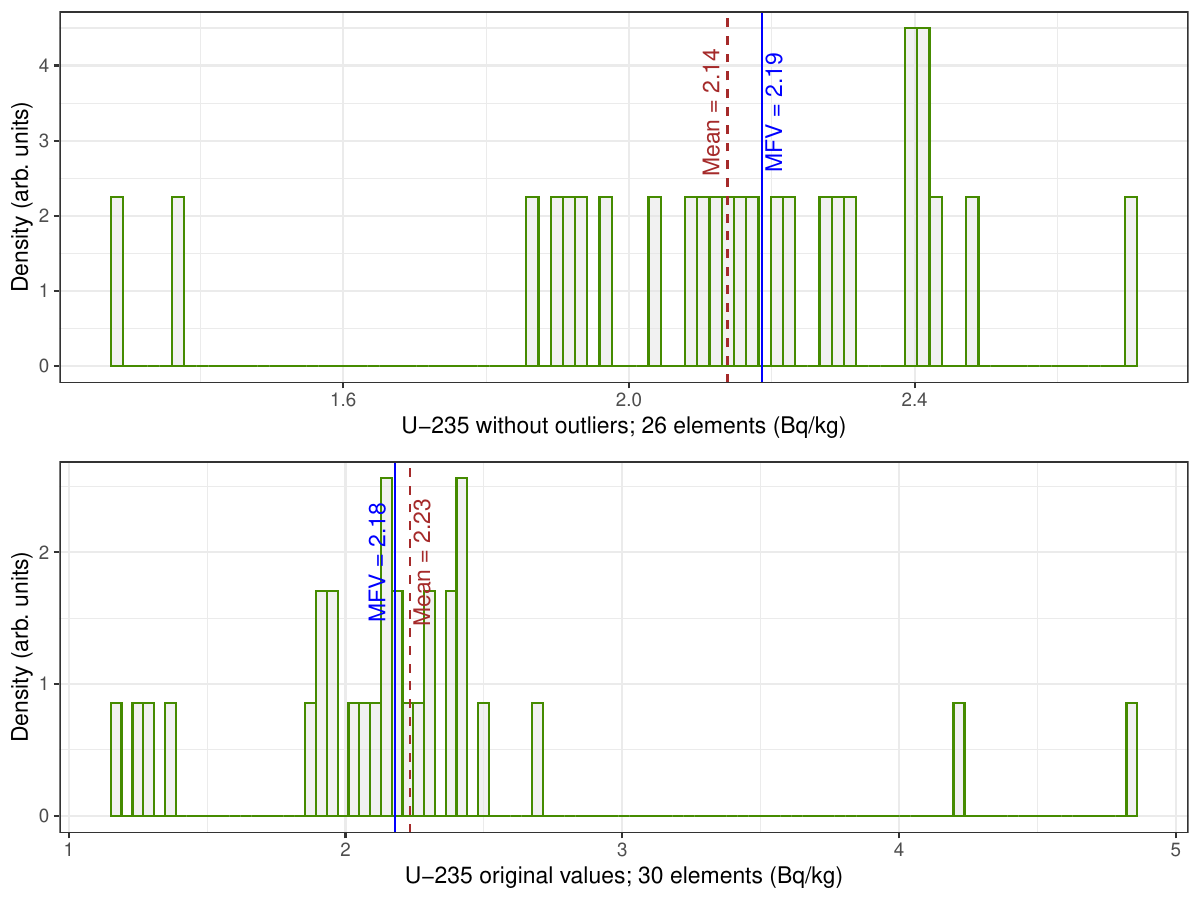}
	\caption{\hl{Histogram} 
		showing the $^{235}$U datasets, including the original dataset with outliers (\textbf{bottom}) and the dataset without outliers (\textbf{top}). The~values are listed in the text. In~each histogram, a~thin solid vertical line represents the MFV for the entire dataset, and~a thin dashed line represents the mean~value. }
	\label{fig:U235his}
\end{figure}

\hl{Figure}~\ref{fig:U235his} presents histograms of the $^{235}$U datasets; the bottom histogram shows the original dataset with outliers, while the top histogram shows the dataset after removing the outliers. For~the dataset without outliers (26 values), the~sample mean and standard deviation are 2.14 and 0.31, respectively. In~comparison, the~MFV for the original dataset is 2.18, while for the dataset without outliers it is 2.19. The~presence of outliers merits further exploration, particularly in data potentially derived from legacy waste, where outliers may indicate contaminants; despite this, the~MFV statistic remain stable, as~the central tendency of the dataset remains nearly the~same.

The Shapiro--Wilk~\cite{shapiro1965analysis} normality test was conducted on the dataset after removing outliers, resulting in a W statistic of 0.91075 and a \emph{p}-value of 0.02743. The~W statistic, which indicates the degree of conformity to a normal distribution, is close to 1, implying a fairly decent fit; however, because~the \emph{p}-value falls below the typical 0.05 threshold, we reject the hypothesis that the data are perfectly normally distributed. This suggests a mild deviation from normality, although~the data still follow a normal distribution. {The Shapiro--Wilk formula and its application are described in further detail elsewhere}~\cite{shapiro1965analysis,das2016brief,zhang2018most}.

{It is important to note that \emph{p}-values, while widely used to measure statistical significance, have inherent limitations}~\cite{amrhein2019scientists}. {Small sample sizes are acceptable when the actual effects being measured are sufficiently substantial to be consistently detectable in such samples}~\cite{greenlandStatisticalTestsValues2016}. {As recommended in the recent statistical literature} (\cite{amrhein2019scientists,dileoStatisticalSignificanceValue2020}), {\emph{p}-values should be interpreted in the context of effect sizes, confidence intervals, and~the broader study~design.}

When evaluating the risk for a nuclear legacy waste dataset that includes outliers, it is recommended to use either the Chebyshev upper confidence limit (UCL) method or nonparametric bootstrapping resampling~\cite{usepaCalculatingUpperConfidence2002}. Bootstrapping~\cite{efron1994introduction} is a strong nonparametric statistical technique for creating approximate confidence limits, {ideally} when the sample data accurately reflect the population. Using the mean statistic with the original dataset results in sensitivity to outliers; thus, the~MFV was employed in this study as the statistical measure in nonparametric bootstrapping, providing a more robust confidence interval for datasets with outliers. In~addition, a~hybrid parametric bootstrapping method is introduced to examine small datasets (fewer than ten elements) selected randomly from the $^{235}$U NORM~dataset.

The small dataset consists of nine measurements of $^{235}$U NORM from the Eastern Desert of Egypt: 1.90 (0.18), 1.87 (0.17), 2.16 (0.20), 2.14 (0.20), 2.31 (0.21), 2.29 (0.21), \mbox{2.70 (0.25)}, 2.08 (0.19), and~2.40 (0.22). To~determine whether these data points are normally distributed, a~Shapiro--Wilk normality test was conducted. The~test results included a W statistic of 0.95553 and a \emph{p}-value of 0.7505. As~this \emph{p}-value is significantly higher than the common threshold of 0.05, the~null hypothesis is not rejected. This result implies that the dataset does not significantly deviate from a normal distribution, which is consistent with the normal distribution~characteristics.

Notably, when applying the Shapiro--Wilk normality test to a small subset of the \(^{235}\text{U}\) NORM activity concentration data, the~results show a normal distribution; however, when the test was conducted on the dataset after removing the outliers, the~results indicate that the data somewhat resemble a normal~distribution.

\vspace{-3pt}
\begin{figure}[H]
	\centering
	\includegraphics[width=\textwidth]{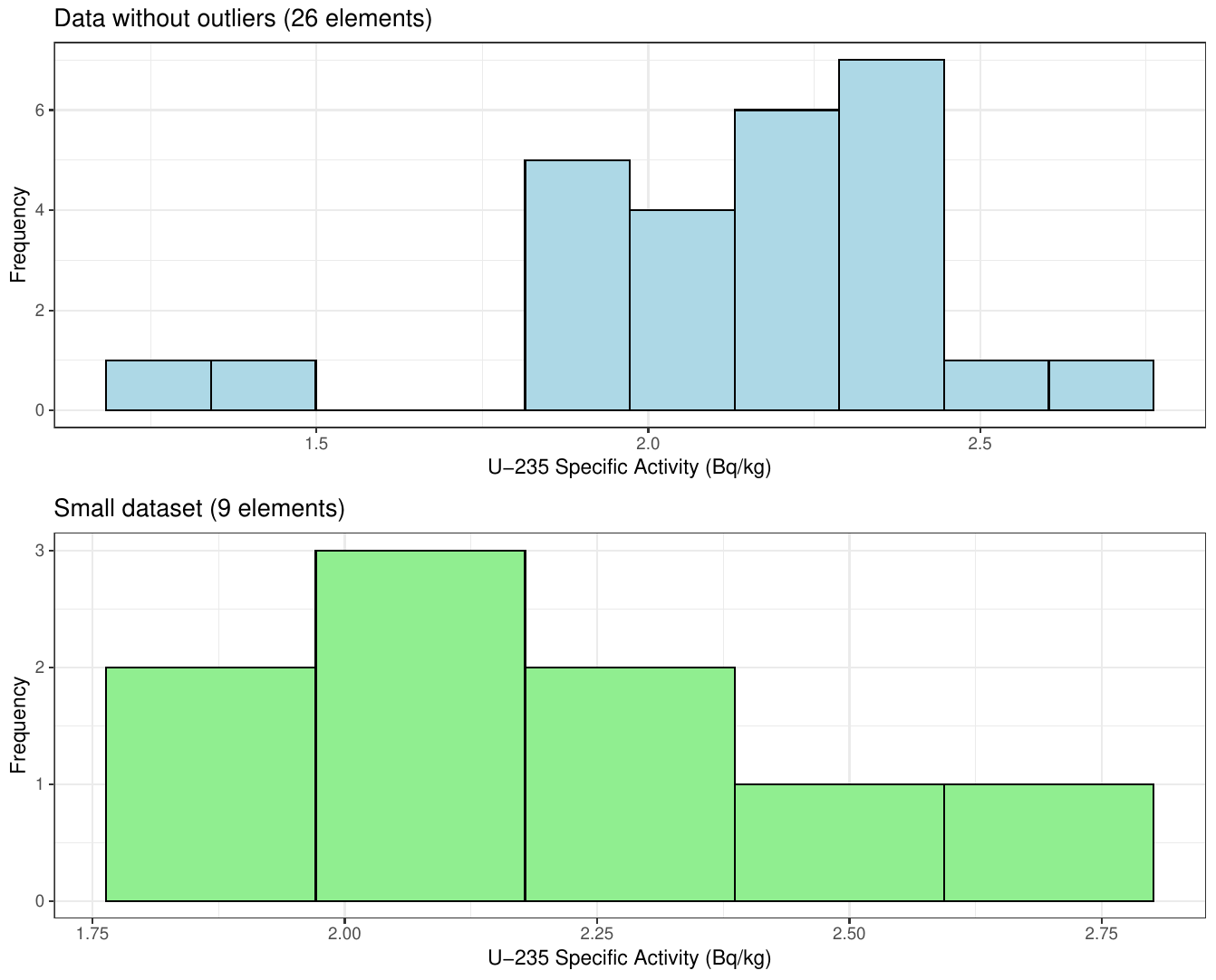} \\
	\caption{\hl{Histograms} 
		showing the $^{235}$U NORM dataset without outliers (26 elements, \textbf{top}) and a smaller subset (9 elements, \textbf{bottom}). The~corresponding values for these datasets are provided in the~text. }
	\label{fig:KolSmirTest}
\end{figure}

After randomly selecting elements for the small subset, we utilized the Kolmogorov--Smirnov test~\cite{an1933sulla,smirnov1948table} to compare the two datasets. As~shown in Figure~\ref{fig:KolSmirTest}, the~$^{235}$U NORM dataset without outliers includes 26 elements, whereas the smaller subset contains 9 elements. We applied the two-sample Kolmogorov--Smirnov test to assess the distribution differences between these two datasets. The~test resulted in a D-statistic of 0.098, indicating the maximum deviation between the cumulative distribution functions. The~\emph{p}-value of 0.9998, which is much higher than the conventional threshold of 0.05, indicates that there is no statistically significant difference between the distributions. The~nearly identical distributions indicates that both datasets likely derive from the same~population.

In this study, we employed a hybrid parametric bootstrapping technique to investigate small datasets containing fewer than ten elements of \(^{235}\)U activity concentrations. Nonparametric bootstrapping traditionally involves resampling from the original dataset without assuming an underlying distribution; however, it neglects the uncertainties inherent in each data point. These limitations make it less effective for datasets with known measurement~uncertainties.

Conversely, traditional parametric bootstrapping is capable of working with small datasets; however, it necessitates prior knowledge of the data's underlying distribution, such as assuming a normal or Gaussian distribution. This assumption becomes problematic when evaluating data from nuclear legacy sites, as~the actual distribution of contaminants is frequently unknown or irregular. Assuming a specific distribution in such cases can result in inaccurate~outcomes.

The HPB method addresses the challenges of data analysis by creating bootstrap samples that capture the uncertainties inherent in each data point without relying on assumptions about a specific population distribution. It goes beyond using the dataset alone by incorporating analytical functions to consider both the data values and their associated uncertainties, thereby providing a more accurate depiction of the variability found in real-world measurements. This makes it particularly effective for estimating confidence intervals and understanding variability within small datasets, which is especially valuable for analyzing nuclear legacy sites. A~detailed procedure for applying the HPB method to a small artificial dataset with only four elements was described in~\cite{golovkoEstimation97RuHalfLife2024}. The~same source also demonstrated the application of HPB to a historical dataset of $^{97}$Ru half-life and the specific activity of $^{39}$Ar derived from underground~measurements.

The upper confidence limit (UCL) based on Chebyshev's inequality is a calculated boundary that indicates where the true mean of a dataset is unlikely to be found. This UCL is derived from the sample mean and sample standard deviation considering the dataset's variability and the sample size. By~choosing a significance level, denoted as \( \alpha \), the~UCL provides a specified confidence level (such as 95\% when \( \alpha = 0.05 \)), ensuring that the true mean does not surpass this upper boundary. This method is beneficial because it does not assume that the data follow a normal distribution, making it ideal for nonparametric datasets or those with~outliers.

The formula for calculating the UCL is
\begin{equation}
	UCL_{1-\alpha} = \overline{x} + \sqrt{\frac{1}{\alpha} - 1} \left( \frac{s}{\sqrt{n}} \right),
	\label{eq:UCL}
\end{equation}
where \( UCL_{1-\alpha} \) represents the one-sided UCL at a confidence level of \( 1 - \alpha \), the~sample mean \( \overline{x} \) is calculated by \( \overline{x} = \frac{1}{n} \sum_{i=1}^{n} x_i \) (where \( x_i \) is each data point in the set), \( n \) stands for the number of sample elements, and~\( s \) is the sample standard deviation, calculated as \( s = \sqrt{\frac{1}{n-1} \sum_{i=1}^{n} (x_i - \overline{x})^2} \). The~significance level \( \alpha \) denotes the desired confidence level (e.g.,~\( \alpha = 0.05 \) corresponds to a 95\% confidence level). This robust method remains effective even with unknown or non-normally distributed data, providing a conservative way of gauging risk or~uncertainty.

\section{
Discussion
}

\begin{figure}[t]
	\includegraphics[width=\textwidth]{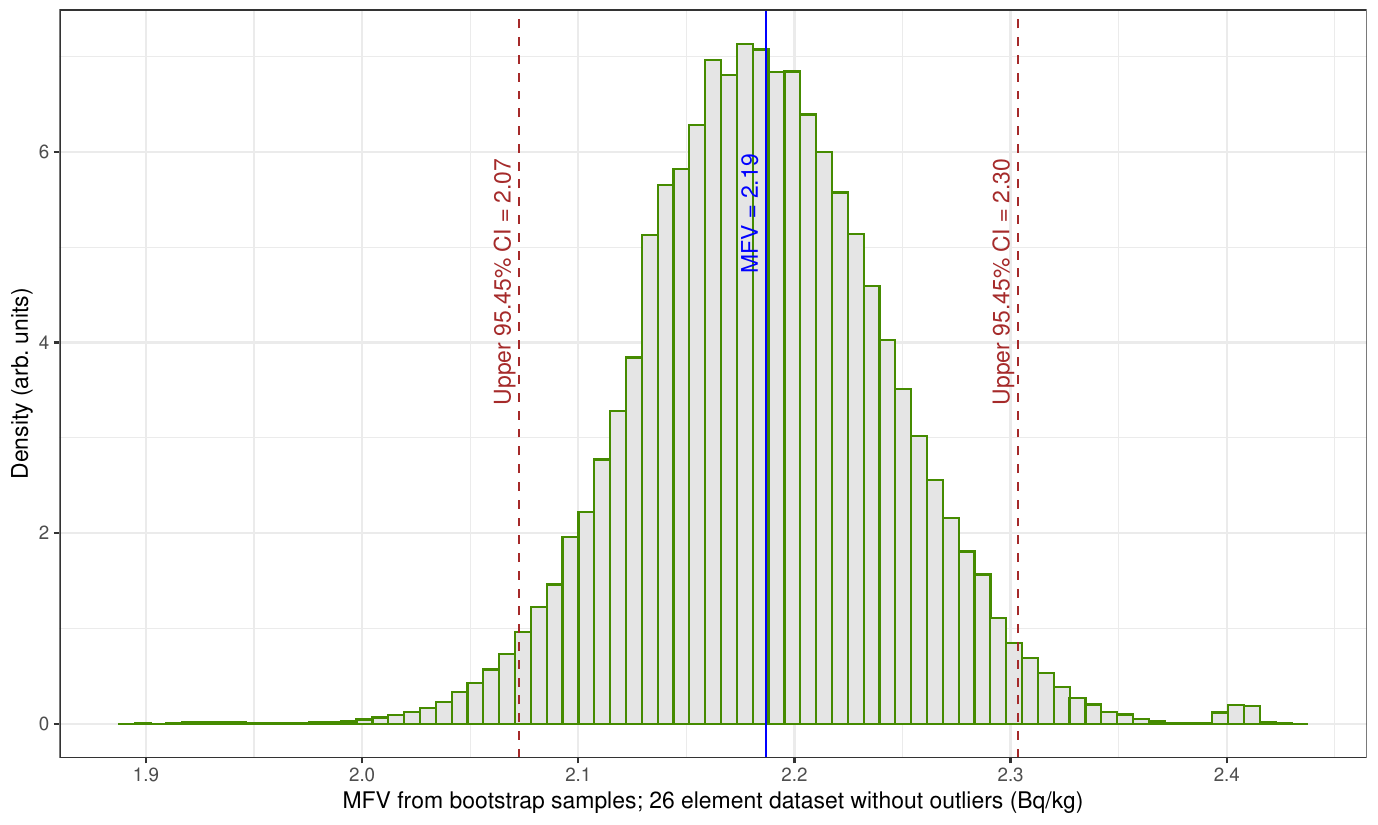}
	\caption{\hl{Histogram} 
 showing the MFV from traditional nonparametric bootstrap samples of the $^{235}$U NORM dataset without outliers (26 elements). In~the histogram, a~thin vertical solid line represents the MFV (2.19), whereas thin dashed lines represent the lower (2.07) and upper (2.30) 95.45\% confidence~limits. }
	\label{fig:MFV-U235bootCI95}
\end{figure}

Figure~\ref{fig:MFV-U235bootCI95} shows the MFV derived from the traditional nonparametric bootstrap samples of the $^{235}$U NORM activity concentration dataset with 26 elements excluding outliers. It also shows the upper 95.45\% confidence limit determined using the percentile method (see~\cite{puth2015variety}). While 3000 bootstrap samples are enough to produce a reliable confidence interval of \mbox{[2.07, 2.30] Bq/kg}, we conducted 210,000 bootstrap samples to ensure consistency with the HPB technique. The~increased number of samples did not alter the confidence interval values. Using the $^{235}$U NORM dataset without outliers and applying Equation~(\ref{eq:UCL}), the~UCL for a significance level \( \alpha = 0.0455 \) corresponds to a 95.45\% confidence level, resulting in 2.42 Bq/kg. This result is not significantly different from the nonparametric bootstrap outcome using the MFV statistic, which is 2.30 Bq/kg.

\vspace{-3pt}

In the analysis of the original $^{235}$U NORM activity concentration dataset consisting of 30 elements, the~UCL calculated using Equation~(\ref{eq:UCL}) for a significance level \( \alpha = 0.0455 \), which aligns with a 95.45\% confidence level, was found to be 2.85 Bq/kg. When applying the Chebyshev inequality for this same confidence level, the~estimates tended to be on the conservative side, often surpassing the 95.45\% mark in their coverage of the sample mean~\cite{singhComputation95Upper2006a}. The~Chebyshev approach does not require assumptions about the data distribution, but~does depend on the known mean and standard deviation of the dataset. The~IQR method indicates that the dataset has four outlying elements. These outliers can skew the mean, which is not robust in such scenarios, and~can inflate the sample standard deviation. Using nonparametric bootstrapping and the MFV statistic, the~95.45\% confidence interval for the original set is [2.05, 2.30] Bq/kg. This interval remains consistent even when the dataset is adjusted to remove the outliers, leaving 26 elements. Both analyses involved 210,000 bootstrap samples, with~the confidence interval estimated using the percentile~method.

\begin{figure}[t]
	\centering
	\includegraphics[width=\textwidth]{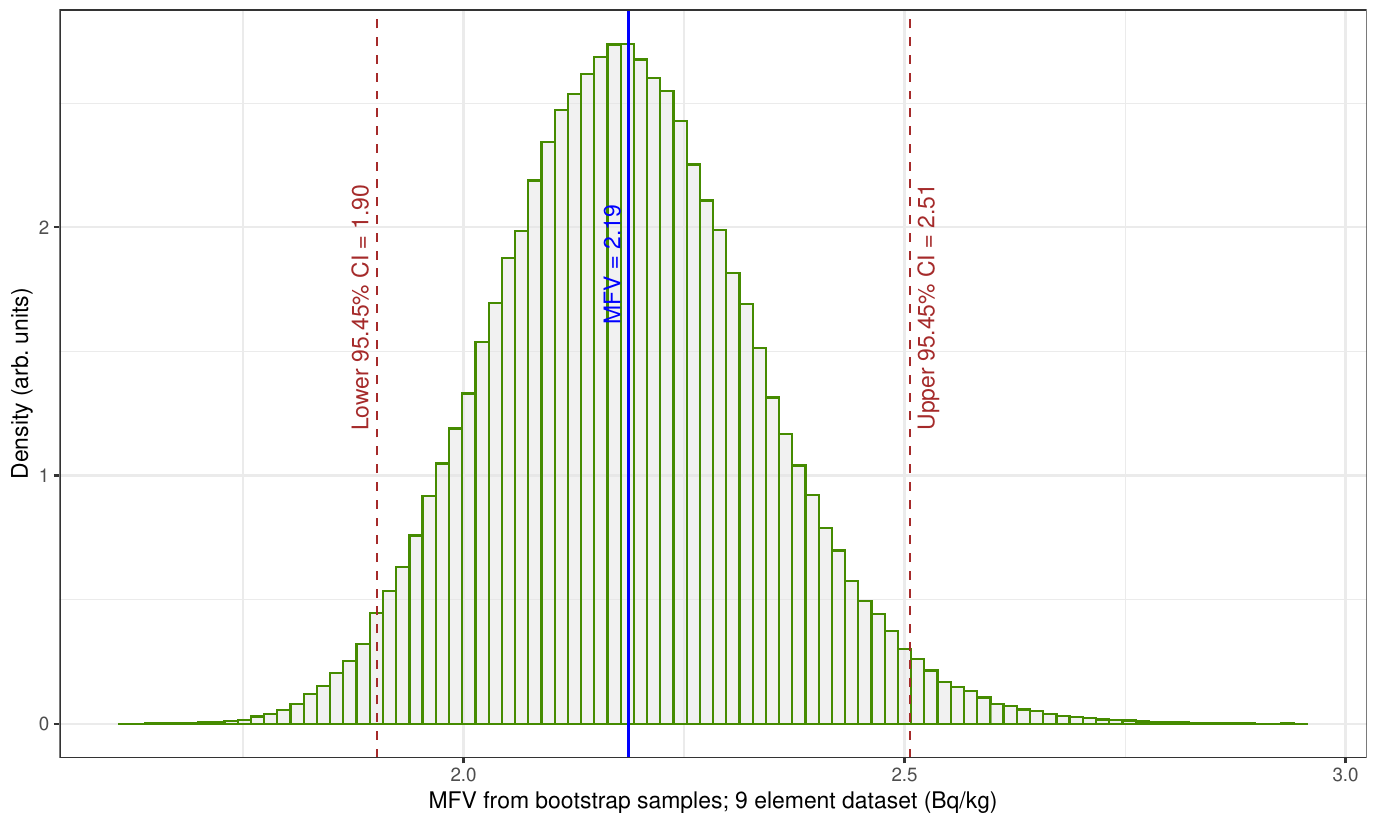} 
	\caption{\hl{Histogram} 
 showing the MFV derived from the HPB samples of the small $^{235}$U NORM dataset (nine elements). The~thin vertical solid line represents the MFV (2.19), whereas the thin dashed lines indicate the lower (1.90) and upper (2.51) 95.45\% confidence~limits. }
	\label{fig:MFV-U235HPBbootCI95}
\end{figure}

The plot in Figure~\ref{fig:MFV-U235HPBbootCI95} presents the MFV obtained through hybrid parametric bootstrapping (HPB) of the small dataset for $^{235}$U NORM activity concentration, consisting of nine elements; also depicted are the upper and lower limits of the 95.45\% confidence interval derived using the percentile method. The~confidence interval at the 95.45\% level was calculated using 210,000 bootstrap samples, spanning from [1.90, 2.51] Bq/kg. When applying the small $^{235}$U NORM activity concentration dataset with nine elements to \mbox{Equation~(\ref{eq:UCL})}, the~UCL for \( \alpha = 0.0455 \), denoting a 95.45\% confidence level, was determined to be \mbox{2.60 Bq/kg}. This result is closely aligned with the HPB outcome from the MFV statistic, showing a 2.51~Bq/kg value, which indicates that the difference is not significant. The~substantial number of bootstrap samples used in the HPB method is crucial, as~it eliminates contributions from the analytical function and aids in considering both the values of the elements and their uncertainties in the relatively small~dataset.

Nonparametric bootstrapping and the UCL based on Chebyshev's inequality overlook the uncertainty in the individual dataset elements. In~order to more conservatively estimate the upper boundary of the concentration, it is possible to use the highest observed value along with its uncertainty at a 95.45\% confidence level, equivalent to 2$\sigma$. Typically, the~concentration data from a site show a skewed distribution, often with some exceedingly high values that indicate localized contamination hot spots that may require~remediation.

\vspace{-3pt}

In this study, some methods may have resulted in very high UCL estimates. According to the risk assessment guidelines cited in~\cite{usepaCalculatingUpperConfidence2002}, when the calculated UCL is much higher than the maximum observed concentration, it is permissible to use this maximum value as the exposure point concentration. However, it is crucial to understand that using the maximum observed concentration may not be a conservative approach, especially with a small dataset, as~in such cases the observed maximum may actually be lower than the true population mean~\cite{usepaCalculatingUpperConfidence2002}.

Criticality safety lessons learned from deactivation and decommissioning (D\&D) efforts~\cite{nirider2003criticality} emphasize the need to use conservative estimates when determining the maximum amount of fissile materials at nuclear legacy contaminated sites during cleanup operations. Poor characterization of facilities planned for D\&D can lead to considerable cost increases and delays in schedules. Thus, it is crucial to perform a detailed risk assessment before starting D\&D projects. Future remediation efforts should include extra caution in nuclear safety evaluations, ensure that they can handle unexpected finds of fissile materials, and~maintain flexibility in~operations.

We establish a conservative approach for estimating the upper confidence limit for the concentrations of fissile materials at legacy nuclear waste sites as follows. First, we estimate the upper limit via the bootstrapping approach combined with the most frequent value. We apply traditional nonparametric bootstrapping when the dataset is sufficient, for~example, when it contains ten or more elements~\cite{golovkoSmartCleanupUsing2025}. For~smaller datasets, we used hybrid parametric bootstrapping. Second, we use the UCL based on Chebyshev's inequality, which applies to datasets of any size. Third, we estimate the upper limit on the basis of the maximum concentration value and twice its uncertainty. Finally, we take the largest value from these methods for further~analysis.

For enhanced understanding, further analysis should involve creating an accurate baseline for fissile materials. This is important because some radioactive nuclei are naturally occurring materials and their concentrations can vary greatly based on location. These materials exist in the Earth's crust and are part of daily human exposure. Thus, it is crucial to start with an assessment of the naturally occurring radioactive material concentrations and/or measurements of the ambient dose rates when addressing nuclear legacy~sites.

Establishing a baseline for the ambient dose can be achieved by using integrating passive detectors such as thermoluminescent dosimeters (TLDs). These devices have been effectively used to measure very low-level ambient dose rates around water shielding at an underground facility, which is essential for ongoing direct dark matter search experiments~\cite{golovkoAmbientDoseDose2023}. The~measurements were initially analyzed using mean statistics, a~traditional averaging method. Additionally, the~same dataset was evaluated with Steiner's MFV statistical technique paired with traditional nonparametric bootstrapping~\cite{golovko2023unveiling}. This combined approach showed that the estimated uncertainty in the ambient dose rates could be reduced by a factor of three. This research underscores the value of using MFV and the nonparametric bootstrapping method to accurately estimate the radiation levels with TLDs. Such precise information is vital for evaluating the potential effects of background radiation in large-scale dark matter experiments conducted in underground~laboratories.

Conservatively estimating fissile material (FM) concentrations helps to determine the mass of FM present in the soil at contaminated sites. This estimation can be accomplished by using the specific activity of a particular FM, which is measured per unit mass or volume and is available for various radioactive substances~\cite{pearceRecommendedNuclearDecay2008}. Additionally, determining the FM mass is essential for assessing the presence of technologically enhanced naturally occurring radioactive materials (TENORMs) in the environment~\cite{ali2019concentrations}. Accurate estimations are crucial because TENORM concentrations can occasionally exceed safe thresholds, potentially leading to environmental and health~hazards.

To demonstrate how to estimate the mass of the fissile materials, this paper adopts a conservative highest estimate for $^{235}$U using the small (nine elements) dataset. Using the HPB statistical method on this dataset provides an estimate of 2.51 Bq/kg with a 95.45\% confidence level. Using Chebyshev's inequality, the~UCL is calculated at 2.60 Bq/kg, also with 95.45\% confidence. The~estimation from the dataset's maximum value (2.40) combined with double its uncertainty (0.22) results in 2.84 Bq/kg; thus, 2.84 Bq/kg is considered the most conservative estimate for the $^{235}$U concentration based on this small dataset. Importantly, this finding is applicable only to this specific dataset. From~the CRL field measurements, we found that any approach in this study could yield a conservative estimate, subject to the data and related~uncertainties.

The granite densities from Egypt's eastern desert were measured, with~results recorded as 2617.07 (8.74), 2590.42 (9.25), 2612.28 (4.72), 2724.35 (3.73), and~2748.23 (3.18) kg/m$^{3}$~\cite{rashwanThermalPhysicomechanicalEvaluation2023}. The~MFV statistical method was employed to determine the central value of these density measurements, resulting in an estimated density of 2614.12 kg/m$^{3}$. Due to the small number of samples, the~confidence intervals were calculated using the HPB method. This provided a 68.27\% confidence interval range of [2602.97, 2737.07] kg/m$^{3}$ and a broader 95.45\% confidence interval of [2587.26, 2750.01] kg/m$^{3}$.

Assuming that the area of potential contamination is 10 m $\times$ 10 m with a depth of 10 m, corresponding to 1000 cubic meters of potentially contaminated soil, the~most conservative estimate for the $^{235}$U concentration based on the previous calculations is 2.84~Bq/kg. To~estimate the mass of $^{235}$U in 1000 cubic meters of granite from the eastern desert of Egypt, the~following steps were performed. First, the~total mass of the granite was calculated using the volume (1000 m$^{3}$) and the MFV-based density estimate of 2614.12~kg/m$^{3}$, resulting in a total mass of 2,614,120 kg. Next, the~total activity of $^{235}$U in this granite was determined using a $^{235}$U concentration of 2.84 Bq/kg, yielding a total activity of 7,424,100.8 Bq. Finally, the~mass of $^{235}$U was derived by converting the total activity using the specific activity of $^{235}$U (79,960(60) Bq/g)~\cite{pearceRecommendedNuclearDecay2008}, resulting in an estimated mass of approximately 92.85 g of $^{235}$U in 1000 cubic meters of~granite.

The Canadian regulatory body describes an ``exempted quantity of fissionable materials'' as being less than 100 g of specific fissile isotopes such as $^{233}$U, $^{235}$U, and~$^{239}$Pu at a licensed site~\cite{canadian2022regdoc}. In~this framework, ``exempted quantity'' means that when the fissionable material is under 100 g, it might not need to adhere to certain regulatory controls or safety standards due to its relatively low potential risk for nuclear~criticality.

In our estimation, 1000\(\,\text{m}^3\) of granite contains 92.85 g of \(^{235}\)U. In~this specific case, the~amount of uranium was exactly 92.85 g, which is below the 100-gram threshold specified in the exemption guidelines. Consequently, under~these regulations the concentration of \(^{235}\)U in this granite volume is under the exempted quantity limit, which means that it may not need the same level of regulatory scrutiny as larger~amounts.

A nonparametric bootstrapping method requires sufficient elements in a dataset to function effectively. When a dataset has fewer than 8 to 10 elements, these methods are not advisable~\cite{singhComputation95Upper2006a}. A~minimum of ten (10) observations is recommended for producing reasonable and reliable estimates, with~more than this number being even better. For~datasets with fewer entries, parametric bootstrapping is suitable under the assumption that there is pre-existing knowledge about the distribution of the dataset elements. In~this work, traditional nonparametric bootstrapping and MFV algorithms have been adapted to assess the upper limit using relatively large datasets (refer also to~\cite{golovkoApplicationMostFrequent2023,golovko2023unveiling,golovkoSmartCleanupUsing2025}). This study shows how hybrid parametric bootstrapping can be used to analyze the upper confidence limit in small datasets (see also~\cite{golovkoEstimation97RuHalfLife2024}).

{As previously mentioned, the~hybrid parametric bootstrapping method is particularly effective for small datasets because it accounts for uncertainties when determining confidence intervals (see also} \cite{golovkoEstimation97RuHalfLife2024}). {To the best of our knowledge, the~HPB method is unique in considering the uncertainty of each individual element in the dataset when estimating a confidence interval. This highlights the importance of proper uncertainty estimation for each~element.
	
HPB provides more cautious estimates of the asymmetric confidence interval because it considers the uncertainty of each data point, unlike the weighted average method, which provides a symmetric confidence interval} \cite{golovkoEstimation97RuHalfLife2024}. {While nonparametric bootstrapping is versatile, it does not consider the uncertainty associated with each individual element in the sample. HPB addresses this limitation by incorporating the uncertainty of each data point. Unlike parametric bootstrapping, HPB does not require knowledge of the exact distribution function of the original dataset; instead, it uses analytical functions to calculate each value in the dataset along with its uncertainty.}

{A good example of how to properly estimate the systematic and statistical uncertainties for the $^{39}$Ar half-life can be found in}~\cite{adhikari2025direct}. {Additionally, a~practical approach to combining statistical and systematic uncertainties into a total error is demonstrated in}~\cite{golovkoApplicationMostFrequent2023}. {Another practical example of uncertainty estimation for $^{39}$Ar specific activity is provided in}~\cite{adhikariPrecisionMeasurementSpecific2023c}. {The $^{39}$Ar specific activity and its associated uncertainty was one of the four elements in the small dataset analyzed using the HPB method for confidence interval estimation}~\cite{golovkoEstimation97RuHalfLife2024}.
{Again, the~HPB method combined with the MFV approach offers a more robust estimation of the confidence intervals. This is particularly valuable for limited datasets, as~it incorporates the uncertainty of each element in the dataset into the analysis.}

\section{
Conclusions
}

This paper has shown that advanced statistical techniques can be successfully adapted to improve the interpretation of sensor data and remediation activities at nuclear legacy sites such as Chalk River Laboratories. This study emphasizes the significance of using various statistical methods to determine the maximum concentrations of contaminants at these locations. Although~specific real-world datasets are excluded here due to security considerations, the~analysis reveals that different statistical methods can produce very conservative upper limit estimates for contaminants. This outcome is crucial for improving nuclear safety~assessments.

When paired with Steiner's most frequent value approach, the~traditional nonparametric bootstrapping method offers strong estimates of data variability and central tendency without depending on the underlying distribution. It is advised that the datasets should have at least ten observations in order to achieve reliable results. However, in~nuclear criticality safety assessments the datasets are often smaller, leading to the adoption of hybrid parametric bootstrapping for these limited data~situations.

It is crucial to understand that bootstrapping does not solve the issue of small sample sizes. It does not create new data or replace any missing information; instead, it provides insights into how additional data samples could behave if taken from a population similar to the initial dataset, which helps in making better decisions for remediation and safety~purposes.

Additionally, using conservative estimates in nuclear safety evaluations is crucial for addressing uncertainties related to contamination. This study emphasizes their importance. On~the other hand, statistical techniques such as the upper confidence limit that rely on Chebyshev's inequality tend to be highly conservative and may overstate risk. Therefore, a~balanced approach tailored to the specific circumstances of the remediation activities should be~employed.

Incorporating naturally occurring radioactive materials (NORM) into baseline assessments is essential because these materials vary greatly between locations. This variability can significantly affect the entire remediation strategy. Understanding the role of NORM, along with that of other contaminants, is vital for creating accurate and effective cleanup~\hl{strategies.} 



\vspace{6pt}
\funding{This research received no external~funding.}

\institutionalreview{Not applicable.}

\informedconsent{\hl{Not applicable.} 
}

\dataavailability{This manuscript has associated data available in the following repository: \href{https://osf.io/3pwmk/}{https://osf.io/3pwmk/}, \hl{(accessed on 11 February 2025)} 
 \cite{golovkoDataOptimizingSensor2024}.}

\acknowledgments{
I am truly grateful for the support and assistance provided by Maria Filimonova. I would also like to extend my appreciation to the management and staff at the Canadian Nuclear Laboratories for fostering an enabling environment for this study, with~special mention of Genevieve Hamilton and David Yuke. The~author would like to thank Stephanie Thomson for her valuable contribution in providing the conceptual site model plot, which was instrumental in supporting the data and analysis presented in this study. Furthermore, I would like to express my gratitude to the anonymous referees for their helpful comments and suggestions, which have greatly contributed to the improvement of this~work.
\hl{During the preparation of this work, the~author used ChatGPT 4.o to check the language of the manuscript. After~using this tool, the~author reviewed and edited the content as needed and takes full responsibility for the content of the~publication.} 
}

\conflictsofinterest{The author declares  that they have no conflicts of~interest.}

\abbreviations{Abbreviations}{
The following abbreviations are used in this manuscript:\\

\noindent
\begin{tabular}{@{}ll}
	CI   & Confidence Interval \\
	CSM  & Conceptual Site Model \\
	CRL  & Chalk River Laboratories \\
	D\&D & Deactivation and Decommissioning \\
	FM   & Fissile Materials \\
	HPB  & Hybrid Parametric Bootstrap \\
	HPGe & High-Purity Germanium \\
	IQR  & Interquartile Range \\
	MFV  & Most Frequent Value \\
	NORM & Naturally-Occurring Radioactive Materials \\
	TENORM & Technologically Enhanced Naturally-Occurring Radioactive Materials \\
	TL   & Thermoluminescent \\
	TLD  & Thermoluminescent Dosimeter \\
	UCL  & Upper Confidence Limit \\
	VOCs & Volatile Organic Compounds \\
\end{tabular}
}


\reftitle{{References} 
}

\end{document}